\documentclass{article}

\usepackage{epsfig}
\usepackage{subfigure}
\usepackage{calc}
\usepackage{graphicx}

\usepackage{amsfonts}
\usepackage{amsmath}
\usepackage{amssymb}
\usepackage{latexsym}

\newcommand{\adots}{\mathinner{\mkern2mu\raise1pt\hbox{.}\mkern2mu%
\raise4pt\hbox{.}\mkern2mu\raise7pt\hbox{.}\mkern1mu}}

\begin{document}

\title{Characterisation of observability and controllability for nonuniformly sampled discrete systems}
\date{}
\author{Amparo F\'{u}ster-Sabater and J.M. Guill\'en\\
{\small Instituto de Electr\'onica de Comunicaciones, C.S.I.C.}\\
{\small Serrano 144, 28006 Madrid, Spain} \\
{\small amparo@iec.csic.es}}

\maketitle

\begin{abstract}

A joint characterisation of the observability and controllability
of a particular kind of discrete system has been developed. The
key idea of the procedure can be reduced to a correct choice of
the sampling sequence. This freedom, owing to the arbitrary choice
of the sampling instants, is used to improve the sensitivity of
system observability and controllability, by exploiting an
adequate geometric structure. Some qualitative examples are
presented for illustrative purposes.

%Keywords: Balancedness, Bit-string model, Combinational generator,
%Design rules

\end{abstract}

\section{Introduction}
\footnotetext{Work supported by Ministerio de Educaci\'{o}n y
Ciencia (Spain).\\
IEE Proceedings, Vol. 135, Pt D, No. 4, pp. 248-252, July 1988.}
The concepts of observability and controllability, introduced
first by Kalman \cite{Kalman}, play an important role in modern
control theory. In fact, these properties often govern the
existence of a solution to an optimal control problem.

Observability and Controllability of discrete-time systems have
also been treated in the literature in a generalised form. A
survey of the main results for discrete system sampled
nonuniformly can be found in \cite{Troch}.

In general, both concepts have been characterised by criteria
mutually independent, but, in this work, it is demonstrated that,
if certain conditions not very restrictive are imposed on the
continuous system, then the characterisations of observability and
controllability for the discrete system can be unified.

There is an important problem which arises in discrete-time
systems but not in continuous-time systems. A linear system which
is completely observable and controllable may lose these
properties after the introduction of sampling. This stresses the
importance of the sampling sequence to guarantee the
above-mentioned internal properties.

There are restrictions on the aperiodic sampling sequence,
stronger than in the periodic case. At any rate, this partial
freedom in the choice of the sampling instants can be conveniently
used to obtain a rather well conditioned system, a reduction in
the propagation of measuring and/or rounding errors etc.

Further interesting applications of these ideas, e.g. the
sensitivity of system observability and controllability, are
treated by exploiting an adequate geometric structure. In this
way, the familiar analytical techniques \cite{Troch} can be
presented in a more intuitive form.

\section{Basic assumptions and general considerations}
This discussion is restricted to:
\begin{description}
\item a) linear time-invariant single-input/single-output
    differential systems of finite order $n$
\item b) continuous-time systems completely controllable and
    observable in the sense given by Kalman \cite{Kalman}
\item c) systems whose transfer function is a strictly proper
    rational function. \end{description}

Consequently, their impulse response $h(t)$ will be a particular
solution of an \textit{n}th-order homogeneous linear differential
equation with constant coefficients of the form

\begin{equation}\label{eq:1}
h^{(n)}(t) + a_1h^{(n-1)}(t)+ \ldots + a_nh(t) = 0 \qquad t \geq 0
\end{equation}

Therefore
\begin{equation}\label{eq:2}
h(t)= \sum\limits_{i=1}^{n} C_i \varphi_i(t)
\end{equation}

where $C_i \in \mathbb{C}$ are constant coefficients. $\varphi_i;
\mathbb{R} \rightarrow \mathbb{C} \;\; (i = 1,\ldots , n)$ is the
fundamental system of solutions of eqn. \ref{eq:1}.

In the state space, the systems considered can be described by the
following equations

\begin{equation*}
\dot{X}(t) = AX(t) + bu(t) \qquad (X_0 = X(0))
\end{equation*}
\begin{equation}\label{eq:3}
Y(t) = c X(t)
\end{equation}

where $X \in \mathbb{R}^n$ denotes the state-vector and $u, y \in
\mathbb{R}$ are the scalar input and output, respectively. The
matrices $A, b, c$ are of appropriate orders and constant.

Canonical realisations in the state space obtained from the
impulse response will be used. In particular,

(i) \textit{Observability canonical form} $(A_{ob}, b_{ob},
c_{ob})$:

$A_{ob}$ is a $n \times n$ bottom-companion matrix with
\begin{equation}\label{eq:4}
(-a_n,-a_{n-1}, \ldots, -a_1)
\end{equation}
in the last row,
\begin{equation}\label{eq:5}
b_{ob} = (h_1, h_2, \ldots, h_n)'
\end{equation}
(' denotes the transpose) where

\begin{equation}\label{eq:6}
h_{i+1} = \frac{d^ih(t)}{dt^i}\bigg |_{t=0} \qquad (i = 0, \ldots, n-1)
\end{equation}

correspond to the \textit{n}-first Markov parameters of the
impulse response $h(t)$ and
\begin{equation}\label{eq:7}
c_{ob}=(1, 0, \ldots, 0)
\end{equation}

(ii) \textit{Controllability canonical form} $(A_{co}, b_{co},
c_{co})$:

\begin{equation}\label{eq:8}
A_{co} = A'_{ob}
\end{equation}
\begin{equation}\label{eq:9}
b_{co} = c'_{ob}
\end{equation}
\begin{equation}\label{eq:10}
c_{co} = b'_{ob}
\end{equation}

All minimal realisations are related by similarity transformations
and, in each particular problem, the most adequate one will be
selected.

\section{Joint characterisation of observability and controllability
of nonuniformly sampled discrete systems}

The observability (controllability) of a discrete system depends
on the observability (controllability) of the continuous-time
system, plus some additional conditions on the sampling sequence.

The problem of observing (controlling) the state of any
realisation by means of the sampling can be reduced to the task of
solving a system of linear equations. Consequently, an adequate
choice of the sampling instants guarantees the compatibility of
this system.

The question of the controllability will be discussed first. The
corresponding results for the observability will be given later.
Finally, a joint characterisation of both internal properties will
be presented.

\subsection{The controllability problem} Let $(A, b, c)$ be an
arbitrary minimal realisation of order $n$ for the kind of systems
under study. The system considered will be completely
\textit{n}-controllable (controllable in $n$ steps)
\cite{Ackermann}, \cite{Troch} if, for any initial state $X_0$ of
eqn. \ref{eq:3}, the system can be directed to $X = 0$ by means of
$n$ impulse inputs applied at $n$ consecutive sampling instants.
The solution of the state-space equation for the system eqn.
\ref{eq:3} at time $t_n$ can be written as
\begin{equation*}
X(t_n)= exp(A\,t_n)X_0 + \sum\limits_{i=0}^{n-1} G_i u_i
\end{equation*}
\begin{equation}\label{eq:11}
= exp(A\,t_n)X_0 + [G_{n-1}, \ldots , G_0]
\left[
  \begin{array}{c}
    u_{n-1} \\
    $\vdots$ \\
    u_0 \\
  \end{array}
\right]
\end{equation}
where
\begin{equation}\label{eq:12}
G_i = exp(A(t_n - t_{i+1})) \int _{-0}^{t_{i+1}}  exp(A(T_{i+1} - r))
\times b\delta (r-t_i)dr \qquad (i = 0, \ldots, n-1)
\end{equation}

\begin{equation}\label{eq:13}
T_{i+1} = t_{i+1}-t_i
\end{equation}

is the length of the sampling interval between two consecutive
sampling instants. The scalar input is

\begin{equation}\label{eq:14}
u(t) = \delta (t-t_i)u_i \qquad (i = 0, \ldots, n-1)
\end{equation}

$u_i$ being the value of the impulse input at the sampling instant
$t_i$. To consider the transference from an initial state to a
final state in $n$ steps, we study the rank of the matrix
$[G_{n-1}, \ldots , G_0]$. Indeed,

\begin{equation}\label{eq:15}
G_i = B \, exp(J(t_n - t_{n-1})) \, exp(J(t_{n-1} - t_i))y_0
\end{equation}

where $J$ is the Jordan canonical form of matrix $A$ and $B$ is
the invertible matrix of the change of basis:

\begin{equation}\label{eq:16}
y_{0} = B^{-1}b
\end{equation}

Therefore, we must compute the value of

\begin{equation}\label{eq:17}
Det = [ y_0, exp(J(t_{n-1} - t_{n-2}))y_0, \ldots, exp(J(t_{n-1} - t_0))y_0]
\end{equation}

or briefly,

\begin{equation}\label{eq:18}
Det = [ exp(J(\alpha_{m})y_0] \qquad  (m = 0, \ldots, n-1)
\end{equation}

with

\begin{equation}\label{eq:19}
\alpha_m = t_{n-1} - t_{n-m-1} \qquad (\alpha_0 = 0)
\end{equation}

By similarity transformations
\begin{equation}\label{eq:20}
y_0 = B^{-1}b = B^{-1}_{co}b_{co}
\end{equation}

where $B_{co}$ is the matrix of the change of basis of $A_{co}$ to
the Jordan canonical form. Then we denote the components of $y_0$
by
\begin{equation}\label{eq:21}
y_0 = (y_{1}^1, \ldots, y_{m1}^1, \ldots, y_{1}^r, \ldots, y_{mr}^r)'
\end{equation}

$m_j \; (j = 1, \ldots , r)$ being the multiplicity of the $r$
different eigenvalues of matrix $A$ with $r \leq n$. It is easy to
see \cite{Kailath} that, for the controllability canonical form,

\begin{equation}\label{eq:22}
B_{co}^{-1} b_{co}= (C_1,C_2, \ldots, C_n)'
\end{equation}

with $(Ci)$ defined as in eqn. 2; thus making use of the Laplace's
expansion by minors, the determinant eqn. 17 can be factorised as
follows:
\begin{equation}\label{eq:23}
Det[\,exp(J \alpha_m)y_0\,]= N_1 N_2 \, Det[\,\varphi_i(\alpha_m)\,]
\end{equation}

\begin{equation}\label{eq:24}
N_1= \frac{1}{0!}\ldots \frac{1}{(m_1-1)!} \ldots \frac{1}{0!}\ldots \frac{1}{(m_r-1)!}
\end{equation}

\begin{equation}\label{eq:25}
N_2=Det \left[%
\begin{array}{ccc}
  \left[%
\begin{array}{ccc}
  y_1^1 & \ldots & y_{m1}^1 \\
  \vdots & \adots & \, \\
  y_{m1}^1 & \, & \, \\
\end{array}%
\right] & \, & \, \\
  \, & \ddots & \, \\
  \, & \, & \left[%
\begin{array}{ccc}
  y_1^r & \ldots & y_{mr}^r \\
  \vdots & \adots & \, \\
  y_{mr}^r & \, & \, \\
\end{array}%
\right] \\
\end{array}%
\right]
\end{equation}

Note that $N_2$ will be non null if, and only if,

\begin{equation}\label{eq:26}
y_{mj}^j \neq 0 \qquad (j = 1, \ldots, r).
\end{equation}

which is guaranteed, according to the previous significance of the
components of $y_0$, because only minimal realisations are
considered.

Finally, $[\varphi_i(\alpha_m)] \; (i = 1, \ldots , n; \; m = 0,
\ldots , n - 1)]$ is an $n \times n$ matrix of the form

\begin{equation}\label{eq:27}
\left[%
\begin{array}{ccc}
  \varphi_1(\alpha_0) & \ldots & \varphi_n(\alpha_0) \\
  \varphi_1(\alpha_1) & \ldots & \varphi_n(\alpha_1) \\
  \vdots & \ldots & \vdots \\
  \varphi_1(\alpha_{n-1}) & \ldots & \varphi_n(\alpha_{n-1}) \\
\end{array}%
\right]
\end{equation}

$(\varphi_i(\alpha))$ being the fundamental system of solutions of
eqn. 1. The value of $Det[\varphi_i(\alpha_m)]$ will depend on the
choice of the sampling instants. Let us now consider one special
aspect here, because it can be the source of some terminological
problems.

The nonsingularity of the matrix $[G_{n-1}, \ldots , G_0]$ is a
necessary and sufficient condition to ensure that any initial
state $X_0$ can be taken to an arbitrary state $X_n$ in $n$ steps:
this is called \textit{n}-reachability. If the state $X_n$
coincides with the origin, then the above condition is sufficient
to insure that any initial state $X_0$ can be taken to the origin
in $n$ steps: this is called \textit{n}-controllability.

The \textit{n}-reachability is a more restrictive condition than
the \textit{n}-controllability. In fact, the
\textit{n}-reachability implies \textit{n}-controllability, but
the converse may not be true.

From eqn. 23 and the previous considerations, the following result
is derived:

\textit{Lemma 1:} The realisation $(A, b, c)$ is completely
\textit{n}-reachable (reachable in $n$ steps) if, and only if, $n$
consecutive sampling instants are chosen in such a way that

\begin{equation}\label{eq:28}
Det[\,\varphi_i(\alpha_m)\,]\neq 0 \qquad (i = 1, \ldots, n;\; \, m = 0, \ldots, n-1)
\end{equation}

\textit{Proof:} The proof is evident from the factorisation of the
expression 23.

Note that, for this kind of system, the \textit{n}-reachability
depends on the characteristic modes of the continuous system and
on the choice of the sampling instants.

We remark that the hypothesis \textit{(b)} in Section 2 is
necessary, because, if this condition is not verified, then the
canonical realizations obtained from the impulse response would
not be actual realisations in the state space for the system
considered. Consequently, the above result would not be true.

In eqn. 14, the scalar input is defined as impulse inputs at the
sampling instants. In practice, a control of the form

\begin{equation}\label{eq:29}
u(t) = u (t_i)=u_i \qquad t_i \leq t < t_{i+1}
\end{equation}

is generally used.

For technical reasons, $u(t)$ must be generated with the aid of a
filter, so that the equations of the device which generates $u(t)$
can be included in eqn. 1. The results are completely analogous.

\subsection{The observability problem}

The observability problem can be discussed without loss of
generality putting $u(t) = 0$. For the same arbitrary minimal
realisation $(A, b, c)$ the observability question is studied. The
system considered will be completely \textit{n}-observable
(observable in $n$ steps) \cite{Ackermann}, \cite{Troch}, if any
initial state $X_0$ of eqn. 3 can be calculated from $n$ values of
the output taken at $n$ consecutive sampling instants.

Therefore, according to eqn. 3 and for the same values $\alpha_0, \alpha_1, \ldots, \alpha_{n-1}$
as before, the following system of linear equations is set:

\begin{equation}\label{eq:30}
 y(\alpha_m)= c \, exp (A\alpha_{m})X_0 \qquad (m = 0, \ldots , n - 1)
\end{equation}

In mathematical terms, the condition of \textit{n}-observability means that

\begin{equation}\label{eq:31}
 rank[c \, exp (A\alpha_{m})]=n \qquad (m = 0, \ldots , n - 1)
\end{equation}

The linear system eqn. 30 can be rewritten as
\begin{equation}\label{eq:32}
 y(\alpha_m)= c B \,exp (J\alpha_{m})z_0 \qquad (m = 0, \ldots , n - 1)
\end{equation}

where $J$ and $B$ are defined as before

\begin{equation}\label{eq:33}
 z_0= B^{-1}X_0
\end{equation}

Now, we study the value of
\begin{equation}\label{eq:34}
 Det[ c B \, exp (J\alpha_{m})] \qquad (m = 0, \ldots , n - 1)
\end{equation}
By similarity transformations

\begin{equation}\label{eq:35}
 cB= c_{ob} B_{ob}
\end{equation}

where $B_{ob}$ is the matrix of
the change of basis of $A_{ob}$ to the Jordan canonical form. It is
known \cite{Kailath} that, for the observability canonical form,

\begin{equation}\label{eq:36}
%\[
\begin{array}{ccccc}

  \, & \, & m_1 & \, & m_r \\
  c_{ob} B_{ob} &= & (\overbrace{1,0,\ldots, 0}, & \ldots, & \overbrace{1,0,\ldots, 0})

\end{array}
%\]
\end{equation}

%\begin{equation}\label{eq:36}
% c_{ob} B_{ob}=(\overbrace{1,0,\ldots, 0}, \ldots, \overbrace{1,0,\ldots, 0})
%\end{equation}
where $m_j \;(j=1, \ldots, r)$ is the multiplicity of the eigenvalues of matrix $A$, with $r \leq n$.

Thus, making use of the Laplace's expansion
by minors, the determinant eqn. 14 can be factorised as
follows:
\begin{equation}\label{eq:37}
 Det[ c B \, exp (J\alpha_{m})] =M_1 M_2 \, Det[ \varphi_i(\alpha_{m})]
\end{equation}

where
\begin{equation}\label{eq:38}
M_1= \frac{1}{0!} \ldots \frac{1}{(m_1-1)!} \ldots \frac{1}{0!} \ldots \frac{1}{(m_r-1)!}
\end{equation}

\begin{equation}\label{eq:39}
M_2= Det \left[
           \begin{array}{cccc}
             I_1 & \, & \, & \, \\
             \, & I_2 & \, & \, \\
             \, & \, & \ddots & \, \\
             \, & \, & \, & I_r \\
           \end{array}
         \right]
\end{equation}

with $I_j \; (j=1, \ldots , r)$
identity matrix of dimension $m_j \times m_j$. Finally, $[\varphi_i(\alpha_{m})] \; (i = 1, \ldots, n), \,(m = 0, \ldots
, n - 1)$ is the same matrix as the one defined in eqn. 27. Now,
from the expression 37, the following result is derived:

\textit{Lemma 2:}
The realisation $(A, b, c)$ is completely \textit{n}-observable (observable in
$n$ steps) if, and only if, $n$ consecutive sampling instants are
chosen in such a way that
\begin{equation}\label{eq:40}
 Det[ \varphi_i(\alpha_{m})] \neq 0 \qquad (i = 1, \ldots, n; \;\, m = 0, \ldots, n-1)
\end{equation}

\textit{Proof:} The proof is evident from the
factorisation of the expression 37.

Note that, for this kind of system, the \textit{n}-observability depends on the characteristic modes of
the continuous system and on the choice of the sampling instants.
We remark that the result is analogous to that of the preceding
subsection.

\subsection{Main result}
The results obtained in the two
preceding subsections can be summarised as follows:

\textit{Theorem:} A
system verifying the conditions \textit{(a)}, \textit{(b)} and \textit{(c)} in Section 2, is
jointly \textit{n}-reachable and \textit{n}-observable, if, and only if, $n$
consecutive sampling instants are chosen in such a way that

\begin{equation}\label{eq:41}
 Det[ \varphi_i(\alpha_{m})] \neq 0 \qquad (i = 1, \ldots, n; \;\, m = 0, \ldots, n-1)
\end{equation}

\textit{Proof:} The
proof is evident from the use of the results obtained in Sections
3.1 and 3.2.

The condition 41 imposes a rather weak restriction
for the choice of the sampling instants. In fact, intervals can be
specified \cite{Fuster}, \cite{Troch} so that complete reachability (observability) is
preserved.

We can also remark that, for this kind of system, the
\textit{n}-reachability and \textit{n}-observability are inseparable concepts. These
systems are either \textit{n}-reachable and \textit{n}-observable or, if not, they
are neither \textit{n}-reachable nor \textit{n}-observable.

It must be noticed that
the condition 41 depends exclusively on the characteristic modes
and the sampling sequence. Thus, the \textit{n}-reachability
(\textit{n}-observability) for these systems is independent of the chosen
realisation.

If the system does not verify the condition \textit{(b)}, then
the results obtained will still be valid for the subsystem
controllable and observable.

Note how reachability involves
statements about inputs and states, while observability about outputs
and states. Nevertheless, both aspects are related to the same
intrinsic system properties. Consequently, the algebraic
characterization for reachability and observability can be
unified.

\section{Problems of sensitivity of system observability and controllability}
In stability theory, it is not enough to know whether a system is
stable or unstable but also the degree of stability. In a similar
way, it is important to know the degree of observability
(controllability) of the system; in the sense that there are
different levels of certitude in the process of resolution of the
corresponding linear equations.

From a heuristic viewpoint, the degree of observability
(controllability) depends on the spatial relationship among the
column vectors of the observability (controllability) matrix. In
particular, maximum observability (controllability) is obtained
when these vectors are mutually orthogonal \cite{Troch}.

Different geometric structures, which show the
evolution in time of the above mentioned vectors, can be found in
\cite{Fuster}. Thus, the question is to choose conveniently the
sampling instants to obtain maximum orthogonality.

We start from
an arbitrary minimal realisation $(A, b, c)$ of the given system
eqn. 3. The controllability matrix

\begin{equation}\label{eq:42}
[G_{n-1}, \ldots , G_0]
\end{equation}

can be easily reduced to the matrix
\begin{equation}\label{eq:43}
[Y_{0},Y_{1}, \ldots , Y_{n-1}]
\end{equation}

\begin{equation}\label{eq:44}
Y_i = exp (J(t_{n-1}- t_{n-1-i}))y_0 \qquad (i = 0, \ldots , n-1)
\end{equation}

($J, \, y_0$ defined as before) more convenient for the
geometric interpretation; as the factor $B\, exp (J(t_n - t_{n-1}))$ in
eqn. 15 affects only the module of the column vectors but not
their spatial position. Analogous results can be obtained making
use of the transpose observability matrix.

Note that matrix 43 depends exclusively on the system
characteristic modes and the sampling sequence; thus the degree of
observability and controllability is independent of the chosen
realisation. Now the problem is reduced to a right choice of the
sampling instants, in such a way that the vectors $Y_i$ are
mutually orthogonal. Some qualitative examples are presented:

\begin{figure}
\begin{center}
\includegraphics[bb= 8 8 150 150, scale=1.2]{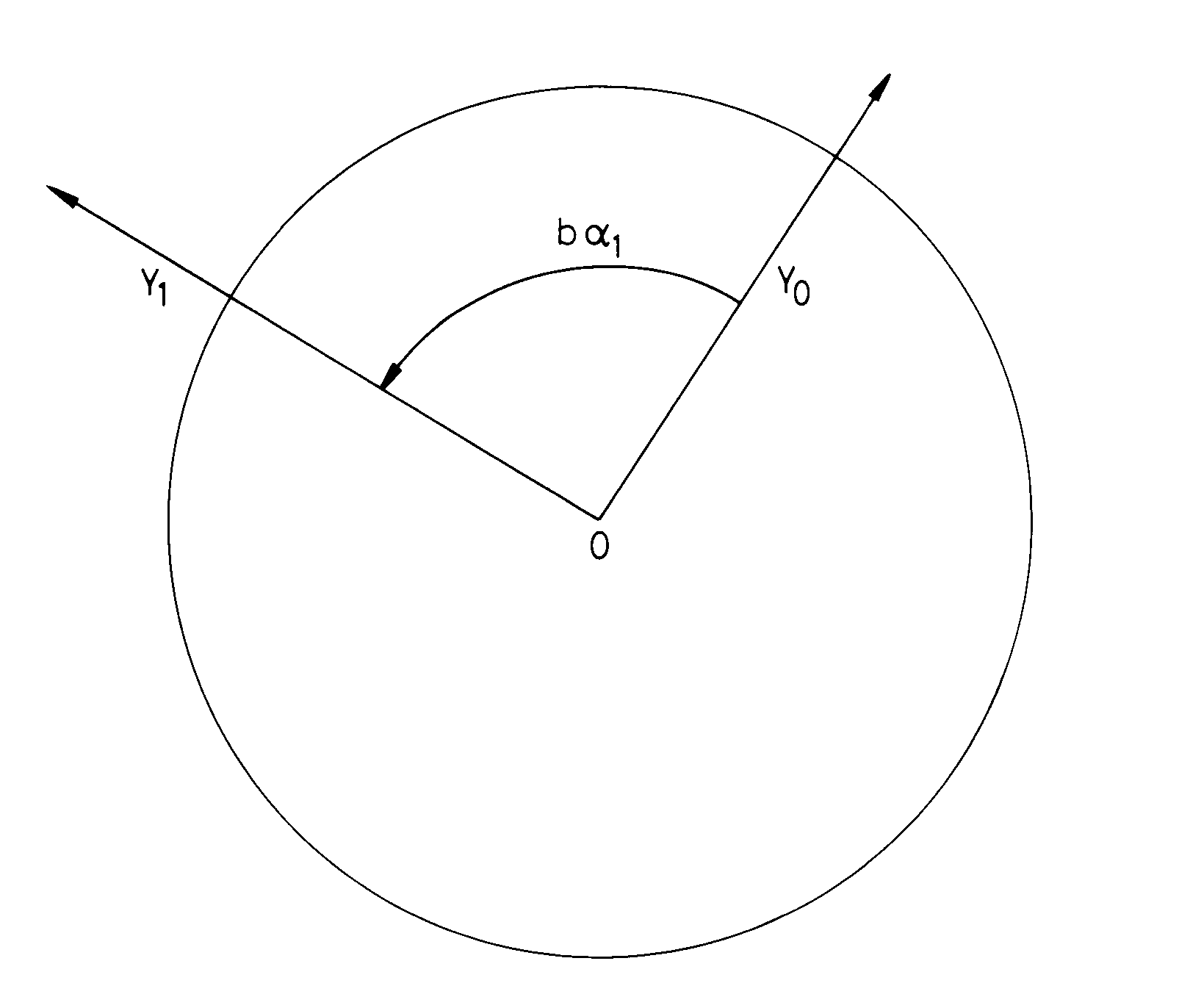}
\caption{Relationships among the vectors $Y_0, Y_1$}
\label{figure:headings1}
\end{center}
\end{figure}

\textit{Example 1:} We are going to
consider a 2nd-order model with a pair of complex eigenvalues
\begin{equation}\label{eq:45}
a + jb \in \mathbb{C} \qquad (b > 0)
\end{equation}
then
\begin{equation}\label{eq:46}
[Y_0, Y_1] = [exp (J\alpha_0)y_0, exp (J\alpha_1)y_0]
\end{equation}

\begin{equation}\label{eq:47}
J = \left[
    \begin{array}{cc}
      a & -b \\
      b & a \\
    \end{array}
  \right]
\qquad (b > 0)
\end{equation}

$\alpha_i, y_0$ defined as before.

As we are in $\mathbb{R}^2$, the geometric
interpretation is very simple. The generic operator $exp (J\alpha)$
applied to the vector $y_0$ can be viewed as follows. It is a
counterclockwise rotation through $b\alpha$ radians, followed by a
stretching (or shrinking) of the length of $y_0$ by a factor $exp (a\alpha)$ \cite{Hirsch}.

From this interpretation, it is easy to see (Fig. 1) that the vectors
$Y_0, Y_1$, will be mutually orthogonal if, and only if,

\begin{equation}\label{eq:48}
b\alpha_1 = b(t_1-t_0)= (2m+1) \frac{\pi}{2} \qquad (m=0, 1, \ldots )
\end{equation}

In this case, maximum observability (controllability) is achieved.
In this situation, expression 48 coincides with the results
obtained by Troch \cite{Troch} for the same model, when the
problem of a minimum transmission of the measuring errors is
discussed.

\textit{Example 2:} Let us now study a
3-order model with a real pole and a complex pair:
\begin{equation}\label{eq:49}
\lambda \in \mathbb{R}, \;\; a + jb \in \mathbb{C} \;\, (b > 0)
\end{equation}

Therefore,
then
\begin{equation}\label{eq:50}
[Y_0, Y_1, Y_2] = [exp (J\alpha_0)y_0, exp (J\alpha_1)y_0, exp (J\alpha_2)y_0]
\end{equation}

with
\begin{equation}\label{eq:51}
J = \left[
      \begin{array}{ccc}
        a & -b & 0 \\
        b & a & 0 \\
        0 & 0 & \lambda \\
      \end{array}
    \right]
\qquad (b>0)
\end{equation}

$\alpha_i, y_0$ as usual. The problem is treated in the 3-dimensional space and the
geometric structure can be described as follows: the generic
vector $Y(\alpha)$ is written as

\begin{equation}\label{eq:52}
Y(\alpha) = (exp (a\alpha)\cos (b\alpha), exp (a\alpha)\sin (b\alpha), exp(\lambda\alpha))'
\end{equation}

thus it is a spiral on the surface revolution:
\begin{equation}\label{eq:53}
z = (x^2 + y^2)^{\lambda / 2a}
\end{equation}
whose form is determined by the real part of
the system eigenvalues.

\begin{figure}
\begin{center}
\includegraphics[bb= 8 8 200 200, scale=1.2]{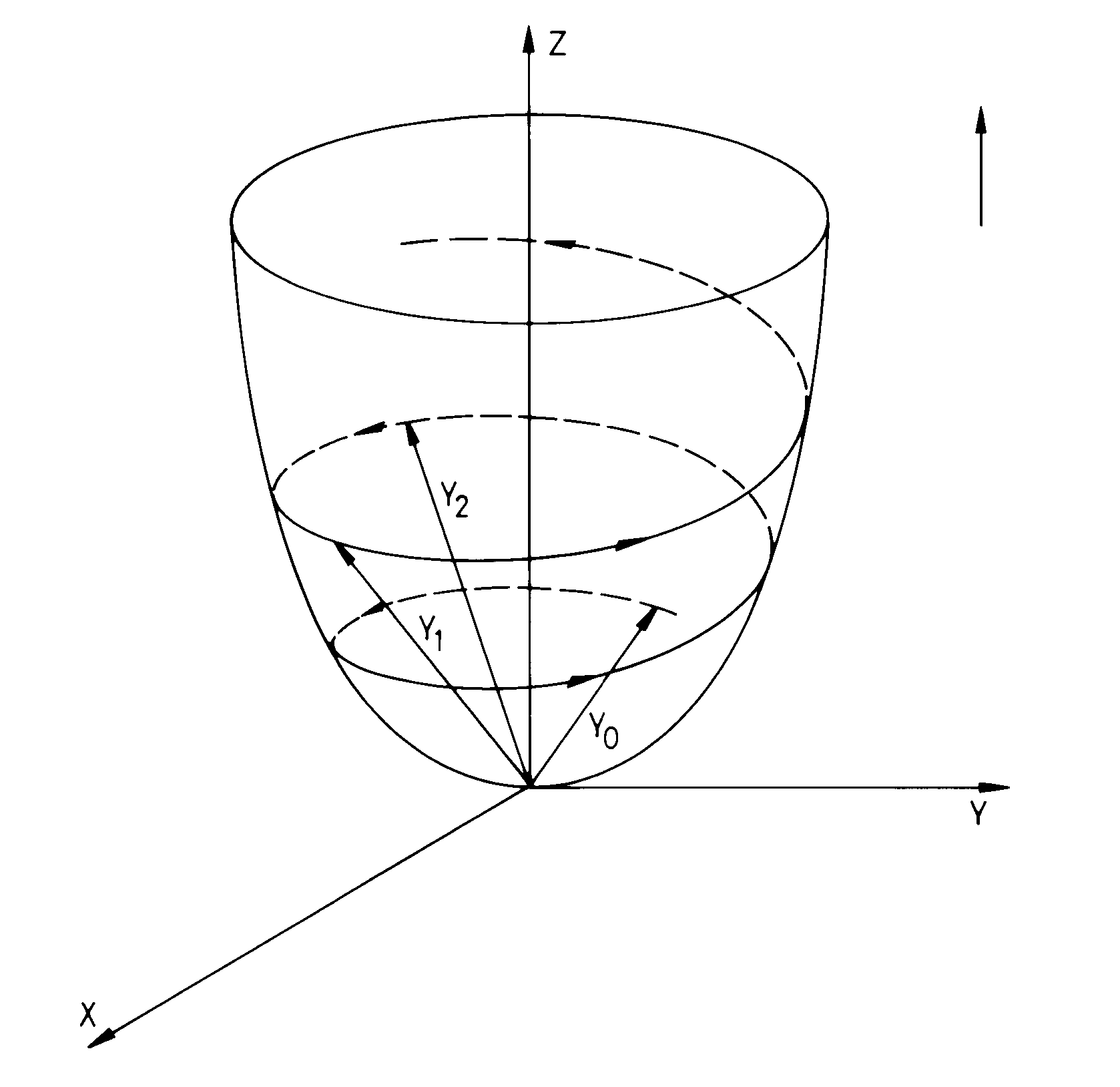}
\caption{Relationships among the vectors $Y_0, Y_1$ and $Y_2$}
\label{figure:headings1}
\end{center}
\end{figure}

The vectors $Y_0, Y_1, Y_2$ have their origin
at the point (0, 0, 0) and their end at the points $Y(\alpha_0), Y(\alpha_1),
Y(\alpha_2)$ of the parametric curve (Fig. 2). Heuristically we can
imagine the same vectors as before (2-dimensional example)
pointing upwards from the \textit{XY}-plane as they have a third component
on the \textit{Z}-axis.

Given $Y_0$ and $Y_1$, the question is now to choose the sampling
instant such that the vector $Y_2$ obeys its motion law and,
simultaneously, is the most orthogonal vector to the other two.

In fact, let $P_0, P_1, P_2$ be the projection on the \textit{XY}-plane of the
vectors $Y_0, Y_1, Y_0 \times Y_1$ , respectively:

\begin{equation}\label{eq:54}
P_0 = (x_0, y_0) = (exp(a\alpha_0)\cos (b\alpha_0), exp(a\alpha_0)\sin (b\alpha_0))
\end{equation}

\begin{equation}\label{eq:55}
P_1 = (x_1, y_1) = (exp(a\alpha_1)\cos (b\alpha_1), exp(a\alpha_1)\sin (b\alpha_1))
\end{equation}

\begin{equation}\label{eq:56}
P_2 = (x_2, y_2) = (y_0 z_1 - y_1 z_0, x_1 z_0 - x_0 z_1)
\end{equation}

with

\begin{equation}\label{eq:57}
z_i = exp(\lambda \alpha_i) \qquad (i=0, 1)
\end{equation}

The sampling sequence must be selected such that
\begin{equation}\label{eq:58}
b\alpha_2 = \arccos \frac{< P_0, P_2 >}{|P_0| \,|P_2|} = M
\end{equation}

then
\begin{equation}\label{eq:59}
b(t_2-t_0)= M + 2\pi m \qquad (m=0, 1, \ldots )
\end{equation}

This number can be understood as the angle of a
counterclockwise rotation. To obtain the more adequate value for
$m$, we analyse the geometry of this situation. Let $Q_2$ be a vector
orthogonal to $Y_0$ and $Y1$ with its end on the revolution surface.
Indeed,
\begin{equation}\label{eq:60}
Q_2 = \mu (Y_0 \times Y_1)
\end{equation}

with $\mu$ easily computable according to eqn. 53; then
the best value of $m$ will be the integer for which the expression

\begin{equation}\label{eq:61}
\Big| (\mu^2(x_2^2 + y_2^2))^{\lambda / 2a} - exp(\frac{\lambda}{b}(M + 2\pi m))\Big|
\end{equation}

is minimum. The process is computed again for each
new sampling instant.

These examples show that the optimal
sampling sequences involve only differences between sampling
instants, but not absolute positions of such instants on the time
axis. It perfectly agrees with the kind of time invariant systems
we are considering.

\section{Conclusions}
It has been shown that, for completely observable and controllable
systems described by eqns. 3, a joint characterisation of the
\textit{n}-reachability and \textit{n}-observability of the
corresponding discrete systems can be given. The formulation
considered stresses the importance of the sampling instants to
preserve the aforementioned internal properties. In this way, such
systems have an additional element for analysis and manipulation.

There are no overrestrictive conditions on the choice of the
sampling sequence. Moreover, it has also been possible to give
strategies for some special cases. At the same time, different
geometric structures have been used to improve the sensitivity of
system observability and controllability, according to a correct
selection of the sampling instants. These geometric structures can
also be used with other performance criterions to give some
insight on the system evolution in terms of familiar concepts.

\end{document}